\documentstyle[aps]{revtex}            
\begin{document}


\draft
\preprint{}
\title{Quaternionic  Electron Theory:\\
       Dirac's Equation}
\author{Stefano De Leo\thanks{{\em E-Mail}: 
              {\tt deleos@le.infn.it} ,    
              {\tt deleo@ime.unicamp.br}    }$^{a,b}$ and   
        Waldyr A.~Rodrigues, Jr.\thanks{{\em E-Mail}:
              {\tt walrod@ime.unicamp.br}    }$^{b}$}
\address{$^{a}$Dipartimento di Fisica, 
                Universit\`a degli Studi Lecce and INFN, 
                Sezione di Lecce\\
                via Arnesano, CP 193, 73100 Lecce, Italia\\
                and\\ 
         $^{b}$Instituto de Matem\'atica, Estat\'{\i}stica e 
                Computa\c{c}\~ao Cient\'{\i}fica, IMECC-UNICAMP\\
                CP 6065, 13081-970, Campinas, S.P., Brasil} 
\date{}
\maketitle


\begin{abstract}

We perform a one-dimensional complexified quaternionic version of the Dirac 
equation based on $i$-complex geometry. The 
problem of the missing complex parameters in Quaternionic Quantum Mechanics 
with $i$-complex geometry is overcome by a nice ``trick''  which allows 
to avoid the Dirac algebra constraints in formulating our relativistic
equation. A brief comparison with other quaternionic formulations is also 
presented.

\end{abstract}


\section{Introduction}
\label{s1}

After the fundamental works of Finkelstein et al.~\cite{FIN} on Quaternionic 
Quantum Mechanics and Gauge Theories, a renewed and increasing interest 
recently appeared~\cite{AD,QM} in the use of noncommutative fields to 
formulate physical theories. In a review paper~\cite{DR}, 
we showed that is possible to give a consistent version of 
Quantum Mechanics by using real and complexified quaternions
as underlying mathematical structure and by adopting a ``complex'' 
geometry~\cite{REM,HB}. We mentioned there, the possibility to 
obtain a {\em natural} formulation of the Dirac equation within a 
complexified quaternionic Quantum Mechanics with $i$-complex geometry.

In the present article, overcoming the problem of the ``apparent'' missing
complex parameters,  we formulate a quaternionic 
version of the Dirac's equation which appears more attractive than the 
previous ones given in the literature~\cite{R,D}. 
Negative energy solutions will 
be quickly obtained from positive energy 
solutions simply by multiplying the latter by the ``complex'' imaginary unit
$\bbox{\iota}$. The {\em spin-flip} will be related to the 
multiplication by the quaternionic imaginary unit $j$. The CPT operation
will map the spinor field $\Psi$ in its dual space: 
Parity and Time reversal will be characterized  
by the ``complex'' involution; Charge conjugation by 
multiplication by ``complex'' ($\bbox{\iota}$) and 
quaternionic ($j$) imaginary units.

The powerful of this new formulation of the Dirac's equation will be also 
evident
when we take its non-relativistic limit. The one-dimensional complexified
quaternionic Dirac equation is obviously not reducible in its dimensions by
performing the non-relativistic limit, contrary to what happens in the real 
quaternionic (2$\rightarrow$1) and complex (4$\rightarrow$2) case. 
\begin{center}
\begin{tabular}{|l|c|c|c|c|} \hline \hline
~{\em Numerical Field} & 
\multicolumn{3}{c|}{\em Equation Matrix Dimensions} & 
~{\em Number of Solutions}~\\ \cline{2-5}
         & ~Dirac~ & ~Schr\"odinger-Pauli~ & ~Schr\"odinger~ & 
~Schr\"odinger~\\ \hline \hline
~Complex  &  4    &          2          &       1       &     1     \\ \hline 
~Real Quaternions    
         &  2    &          1          &       1       &     2     \\ \hline 
~Complexified Quaternions
         &  1    &          1          &       1       &     4     \\ \hline 
\hline
\end{tabular}
\end{center}
In discussing the non-relativistic Schr\"odinger equation we find in 
its real quaternionic formulation a 
{\em belated theoretical discovery of spin}~\cite{PRD}, 
working with complexified quaternions we like talking of  
{\em belated theoretical discovery of positron}. It is worth to mention that 
the non-relativistic Schr\"odinger approximation to Dirac's equation formulated
with the Clifford algebra, $Cl_{1,3}$, show also that spin is present in
Schr\"odinger theory, but it is ``frozen''.

In the literature we find two different quaternionic formulations of the 
Dirac equation, with complex geometry, which reproduce the standard results. 
The first one, performed 
in 1989~\cite{R}, is obtained by $2\times 2$ real quaternionic 
matrices, 
the second one, dated 1996~\cite{D}, overcomes previous 
difficulties, i.e. 
non physical doubling of solutions~\cite{DS,DS1}, 
and allows a one-dimensional complexified
quaternionic  representation of the Dirac algebra and 
consequently a one-dimensional 
version of the Dirac 
equation. These formulations, notwithstanding the reduced dimensions of the
spinors, reproduce the standard results thanks to the  doubling 
(real quaternions) and quadrupling (complexified quaternions) of solutions
due to complex geometries~\cite{DR}. Nevertheless, we do not have serious 
reasons in  
preferring quaternionic to complex formulations. The only apparent advantage in
using complexified quaternions is given by the possibility to translate back
the one-dimensional complexified quaternionic Dirac's equation in a {\em new} 
equivalent complex equation, performed by the Pauli's algebra and so by 
$2\times 2$ complex matrices. At first glance, this would appear very strange 
because of the 4-dimensionality requested for the $\gamma$-matrices.  
Nevertheless, we can re-obtain the right complex parameters counting by allowing
a left/right action of two-dimensional matrices. 
By passing from complex to complexified quaternions we show that the 
standard Dirac equation, written in the standard formalism by using 
the Clifford 
algebra $Cl_{4,1}$, can be rewritten by using the Clifford algebra
$Cl_{3,0}$\cite{Z}, called the Pauli algebra.  In spite
of this, we must admit a {\em not} elegant version of the Dirac equation by 
complexified quaternions and $\bbox{\iota}$-complex geometry~\cite{D}.  
Our aim in this paper is to present a complexified quaternionic formulation
of the Dirac equation where the non-commutativity of the quaternionic field
represents an {\em advantage} and {\em not} an undesired and useless 
complication.

This work is structured as follows: After a mathematical introduction to the 
complexified quaternionic algebra, sec.~\ref{s2}, we briefly recall the 
quaternionic formulations of the electron theory found in literature, 
sec.~\ref{s3}. The {\em new} complexified quaternionic version of the Dirac 
equation is given in sec.~\ref{s4}. We discuss the CPT operation in 
sec.~\ref{s5} and draft our conclusions in the last section.


\section{Complexified Quaternionic Algebra}
\label{s2}

In this section, we introduce the complexified quaternionic algebra and the 
so-called ``barred'' operators. For a complete review of the quaternionic 
mathematical language used in this paper the reader can consult 
ref.~\cite{DR}. 

The complexified quaternionic
algebra is a quaternionic algebra, ${\cal H}(1,\vec{h})$, over a complex 
field, ${\cal C}(1,\bbox{\iota})$,
\begin{equation}
{\cal H}_c = \left\{ c_0 + \vec{h} \cdot \vec{c}~,~~~\vec{h}\equiv(i,j,k)~,
                   ~~~\vec{c}\equiv(c_1,c_2,c_3)~,
                   ~~~~~~~c_{0,1,2,3} \in {\cal C}(1,\bbox{\iota})
           \right\}~,
\end{equation}
with operation of multiplication defined according to the following rules 
for the imaginary units
\begin{eqnarray*}
\bbox{\iota}^2                          & = & -1~,\\ 
i^2=j^2=k^2                             & = & -1~,\\
ijk                                     & = & -1~,\\
\left[ \, \bbox{\iota} \, , \, \vec{h} \, \right]  & = &  0~.
\end{eqnarray*}
Working with complexified quaternions we have three different (independent)
opportunities to define conjugation operations
\begin{eqnarray*}
q_c^{\bullet} & = & c_0^* + \vec{h} \cdot \vec{c}^{\, *}~,\\
q_c^{\star} & = & c_0 - \vec{h} \cdot \vec{c} ~,\\
q_c^{\dag} & = & c_0^* - \vec{h} \cdot \vec{c}^{\, *}~,
\end{eqnarray*}
where $*$ indicates the standard complex conjugation 
($\bbox{\iota} \rightarrow - \bbox{\iota}$). Note that 
$q_c^{\dag}=q_c^{\bullet \star}= q_c^{\star \bullet}$. The $\bullet$ involution
is an auto-morphism, $(q_c p_c)^{\bullet}=q_c^{\bullet} p_c^{\bullet}$, while
the $\star$ and $\dag$ conjugations are anti-auto-morphisms, that is 
$(q_c p_c)^{\star} = p_c^{\star} q_c^{\star}$ and
$(q_c p_c)^{\dag} = p_c^{\dag} q_c^{\dag}$.

Due to the non-commutative nature of the quaternionic multiplication, we must
distinguish between the left and right-action of our imaginary units 
$i$, $j$, $k$. We introduce {\em barred operators} to represent the right
action of the three quaternionic imaginary units. Explicitly,
\[ 1\mid i \, , ~1\mid j \, , ~1\mid k \]
will identify the right-multiplication of $i$, $j$, $k$ and so
\[ \left( 1 \mid \vec{h} \right) \, q_c \equiv q_c \, \vec{h} ~ . \]

In this formalism, the most general transformation on complexified quaternions
will be given by
\begin{equation} 
\label{fbo}
q_c + p_c \mid i + r_c \mid j + s_c \mid k ~~~~~~~   
q_c,p_c,r_c,s_c \in {\cal H}_c ~.
\end{equation}
Such an object represents an $\bbox{\iota}$-complex linear (complexified
quaternionic) operator, characterized by 16 $\bbox{\iota}$-complex parameters.
Obviously, we can also require $i$-complex linearity for our transformations. 
In this case the most general ($i$-complex linear) transformation which can be
performed on complexified quaternions will be characterized by ``only'' 8
$i$-complex parameters
\begin{equation} 
\label{pbo}
q_c + p_c \mid i~.
\end{equation}
Going back to real quaternions, because of the missing imaginary complex 
unit $\bbox{\iota}$, we can define only $i$-complex linear operators
\begin{equation} 
\label{rbo}
q + p \mid i~~~~~~~q,p \in {\cal H}~,
\end{equation}
characterized by 4 $i$-complex parameters. Why this counting of ``complex''
parameters? Why ``complex'' geometry?

We showed in previous papers~\cite{DR,PAP}  
that the choice of a complex projection of quaternionic inner products, 
also called complex geometry~\cite{REM}, gives the possibility to formulate a 
consistent quaternionic 
version of standard (complex) Quantum Mechanics. Many difficulties, due to
the non-commutative nature of quaternionic multiplication, are soon overcome.
See for example the definition of an appropriate momentum operator~\cite{DR}. 
The choice of a complex geometry implies:

1 - The introduction of ``new'' anti-hermitian imaginary units
\begin{center}
\begin{tabular}{cr}
$\left( 1 \mid \vec{h} \right)^{\dag} = -  1 \mid \vec{h}$
& ~~~~~~~$\bbox{\iota}$-complex geometry~,\\
 & \\
$\left( 1 \mid i \right)^{\dag} = -  1 \mid i$
& ~~~~~~~$i$-complex geometry~.
\end{tabular}
\end{center}

2 - Quadrupling of solutions for complexified quaternions
\begin{center}
\begin{tabular}{lr}
$1 \, , ~ i \, , ~ j \, , ~ k$
& ~~~~~~~$\bbox{\iota}$-complex geometry~,\\
 & \\
$1 \, , ~ j \, , ~ \bbox{\iota} \, , ~ \bbox{\iota} j $
& ~~~~~~~$i$-complex geometry~,
\end{tabular}
\end{center}
and doubling of solutions for real quaternions
\begin{center}
\begin{tabular}{lr}
$1 \, , ~ j $
& ~~~~~~~$i$-complex geometry~.
\end{tabular}
\end{center}

The previous counting of ``complex'' parameters suggests to relate  
{\em barred} operators and quaternionic field to 
complex matrices and column vectors in the following way,
\begin{center}
\begin{tabular}{lccc}
Complexified Quaternions~:~~~~~~~~ &
$q_c + p_c \mid i + r_c \mid j + s_c \mid k $
& $~~\leftrightarrow ~~$ &
$4\times 4$ complex matrices~, \\
 &  & & \\
 & $c_0 + \vec{h} \cdot \vec{c} $
& $~~\leftrightarrow ~~$ &
$\left( \begin{array}{c} c_0\\ c_1 \\ c_2 \\ c_3 \end{array} \right)$~,\\
 & & & \\
Real Quaternions~: & $q + p \mid i $
& $~~\leftrightarrow ~~$ &
$2\times 2$ complex matrices~, \\
 & & & \\
 & $z + j \tilde{z} $
& $~~\leftrightarrow ~~$ &
$\left( \begin{array}{c} z\\ \tilde{z} \end{array} \right)$~.
\end{tabular}
\end{center}
This allows one-dimensional complexified quaternionic~\cite{D} and 
two-dimensional
real quaternionic~\cite{R} versions of the Dirac equation. 
We also have the necessary tools to performing a set of translation 
rules for passing back and forth between standard (complex) and quaternionic 
Quantum Mechanics~\cite{DR,T}.
Note that when working with complexified quaternions this is 
achieved by adopting 
$\bbox{\iota}$-complex geometries. Up to now, the use of $i$-complex 
geometries seemed to be avoided because of the missing complex parameters 
in the barred operators structure~(\ref{pbo}).


\section{Quaternionic Dirac's Equation: A brief review}
\label{s3}

In this section we briefly recall the formulation of the Dirac's equation by 
real and complexified quaternions.

\subsection{Real Quaternionic Version: The Milestone}
\label{s31}

In 1989~\cite{R}, Rotelli derived a quaternionic version of the free 
particle Dirac's equation, which required, for its development, the use of the 
complex scalar product. He observed that the need to use the complex 
scalar product no longer relies solely on arguments relative to tensor 
product space (multi-particle systems)~\cite{HB} but is explicit in the single 
free 
particle wave function.

The first important modification, that must be made is, the rewriting the 
standard Dirac equation
\[ i\partial_t \psi = \left( \vec{\alpha} \cdot \vec{p} + \beta m \right) 
\psi~,\]
where $\psi\equiv \psi(x)$ are $4\times 1$ complex matrices, in the form
\[ \partial_t \psi i = \left( \vec{\alpha} \cdot \vec{p} + \beta m \right) 
\psi~,\]
where now $\psi\equiv \psi(x)$ stands for real quaternionic column vectors.
The right position of the imaginary unit $i$ 
guarantees the norm conservation of $\psi$
\[ \partial_t \int d\tau \psi^{\dag} \psi =
   i \int d\tau \psi^{\dag} H \psi - 
   \int d\tau \psi^{\dag} H \psi i ~,\]
since ``$\int d\tau \psi^{\dag} H \psi$'' is real and hence commutes with $i$.

The relativistic covariance is obtained by redefining the action of the 
momentum operator $\vec{p}$ as follows
\[ \vec{p} \psi \equiv - \vec{\partial} \psi i~.\]
The hermiticity of $\vec{p}$ imposes the choice of the 
{\em complex scalar product}~\cite{R}.

The $\gamma$-matrices can be now expressed by $2\times 2$ real quaternionic 
matrices
\[ 
\gamma_0 = 
           \left( \begin{array}{cc} 1 & 0\\ 0 & $-$1 \end{array} 
           \right)~,~~~~~~~
\vec{\gamma} = \vec{h}  
           \left( \begin{array}{cc} 0 & 1\\ 1 & 0 \end{array} \right)~,
\]
and the solutions read
\begin{eqnarray*}
E>0~~~~~~~~~~ & N \left( \begin{array}{c} 1 \\ 
                 $-$ \frac{\vec{h} \cdot \vec{p}}{|E|+m} \end{array}
           \right) e^{-ipx}~,~~~~ 
            & N \left( \begin{array}{c} 1 \\ 
                 $-$ \frac{\vec{h} \cdot \vec{p}}{|E|+m} \end{array}
           \right) j e^{-ipx}~,~~~\\
E<0~~~~~~~~~~ & N \left( \begin{array}{c} 
                 $-$ \frac{\vec{h} \cdot \vec{p}}{|E|+m} \\ 1  \end{array}
           \right) e^{-ipx}~,~~~~
            & N \left( \begin{array}{c}
                 $-$ \frac{\vec{h} \cdot \vec{p}}{|E|+m} \\ 1 \end{array}
           \right) j e^{-ipx}~,~~~~
\end{eqnarray*}
where 
\[ N = \sqrt{\frac{|E|+m}{2}}~. \]
In such a formalism the multiplication by the quaternionic imaginary unit $j$
gives a {\em spin-flip} and this implies the desired doubling of solutions in
the quaternionic version of the Schr\"odinger's equation. The so-called
{\em belated theoretical discovery of spin}~\cite{PRD}. Inspired by the 
Rotelli's work, we find in the literature many papers based on real 
quaternionic Quantum Mechanics with $i$-complex geometry. Among these, we cite
the quaternionic versions of the Lagrangian formalism~\cite{LF}, Electroweak 
Model~\cite{EM} and Grand Unification Theories~\cite{GUT}.

\subsection{Complexified Quaternionic Version: Hope and Disappointment}
\label{s32}

Various formulations of Dirac equation on the complexified field were 
considered since the 1930's. A pioneer in this field was certainly 
Conway~\cite{CON}; more recent presentations can be found in the 
Edmonds paper~\cite{DS} and Gough~\cite{DS1}. When written in this manner, a 
doubling of solutions from four to eight occurs. The possible physical 
significance of these additional solutions has been a matter of 
speculation~\cite{DS2}.

In a recent article~\cite{D}, it was showed that such a doubling of solutions
is strictly connected with the use of reducible matrices and so there is
{\em no new physics} in the quaternionic Dirac equation. Indeed, 
by following the standard Dirac approach it is possible to formulate a
one-component equation with only four solutions~\cite{D}. 
The previous ``unphysical''
doubling of solutions is overcome by allowing a one-dimensional representation
for the $\gamma$-matrices by barred complexified quaternionic operators
\[ q_c + p_c \mid i + r_c \mid j + s_c \mid k~. \]
Nevertheless, we must admit that such a version of the Dirac's equation is 
neither elegant nor simple. In few words, it appears {\em unnatural}:
Complicated spinors structures, unclear CPT interpretation, etc. We do not 
have any 
particular reason to prefer this version to the complex formulation.
The other possibility, that is to performing a complexified quaternionic 
version of the Dirac equation by using $i$-complex geometry, appears 
unlikely, due
to the {\em missing} complex parameters within $i$-complex linear barred
operators.

We conclude this section by discussing the possibility to write down a Dirac
equation based on the Clifford algebra $Cl_{3,0}$. The formulation of 
Dirac's theory by complexified quaternions, implies the possibility to rewrite
the Dirac equation by the Pauli's algebra.  
The matrices $\vec{\sigma}$ generate the algebra of $2\times 2$ 
matrices with complex numbers as entries $M_2({\cal C})$. The matrix algebra
$M_2({\cal C})$ has the following basis over $\cal R$
\begin{center}
\begin{tabular}{|c|c|} \hline \hline
~~~~~~~~~~$M_2({\cal C})$~~~~~~~~~~ & ~~~~~~~~~~${\cal H}_c$~~~~~~~~~~\\ \hline
$\openone$  &          
$ 1$             \\ 
$\sigma_1$, $\sigma_2$, $\sigma_3$  & 
$\bbox{\iota} i$, $\bbox{\iota} j$, $\bbox{\iota} k$\\
$\sigma_3 \sigma_2$, $\sigma_1 \sigma_3$, $\sigma_2 \sigma_1$ 
 & 
$i$, $j$, $k$\\
$\sigma_1 \sigma_2 \sigma_3$ &  
$\bbox{\iota}$\\ \hline \hline
\end{tabular}
\end{center}
The above table also gives the corresponding basis of the complexified 
quaternionic algebra.

By translation from our complexified quaternionic version we can obtain 
a formulation of the Dirac equation by $M_2({\cal C})$. We identify 
the spinor fields by $2\times 2$ complex 
matrices and  obtain the needed complex freedom degrees.
The most general transformation on the $4$-dimensional complex vector
column
\begin{eqnarray}
\label{vc}
\left( \begin{array}{c} 
       \psi_1 \\ \psi_2 \\ \psi_3 \\ \psi_4 
       \end{array}
\right)
\end{eqnarray}
is obviously performed by $4\times 4$ matrices, 16 complex parameters. By
rewriting  the previous 4-dimensional vector column by a $2\times 2$ 
complex matrix
\begin{eqnarray}
\label{m2}
\left( \begin{array}{cc}
       \psi_a & \psi_b \\ \psi_c & \psi_d
       \end{array}
\right)~,~~~~~~~\psi_a=\psi_1-i\psi_4~,~\psi_b=-\psi_3+i\psi_2~,
               ~\psi_c=\psi_3-i\psi_2,~\psi_d=\psi_1+i\psi_4~,
\end{eqnarray}
we find again 16 complex parameters within the most general transformation
on our ``new'' spinors. Indeed, by allowing left/right action for the Pauli's 
matrices, we have
\[ M_0 + M_1 \mid \sigma_1 + M_2 \mid \sigma_2 + M_3 \mid \sigma_3~,\]
where $M_{0,1,2,3}$ are $2\times 2$ complex matrices, 
and so we restore the 16 complex parameters characterizing the 
standard action on spinor fields.


\section{Quaternionic Dirac's Equation: Its natural formulation}
\label{s4}

Let us work within complexified quaternionic Quantum Mechanic with 
$i$-complex geometry.  In finding the representation of gamma matrices 
satisfying the Dirac's algebra, 
we have no problems with the $\vec{\gamma}$-matrices, in fact we immediately 
find as suitable choice
\[ \vec{\gamma} = \vec{h} \equiv (i,j,k),~~~
\{ \, h^m, \, h^n \, \} = 2 g^{mn}~~~(m,n=1,2,3),~~~
\vec{h}^{\, \dag} = - \vec{h}~.
\]
Nevertheless, we cannot find a quaternionic number which anti-commutes 
with $\vec{h}$, and consequently we cannot give a (complexified) quaternionic 
representation for the $\gamma^0$-matrix. Working in complexified 
quaternionic QM with $\bbox{\iota}$-complex geometry, the problem is overcome 
by using
two {\em different} barred quaternionic imaginary units in representing 
$\gamma^0$ and $\vec{\gamma}$. Explicitly
\[ 
\gamma^{0} = i\mid i ~~~~~\mbox{and}~~~~~
\vec{\gamma} = \bbox{\iota} \vec{h} \mid j~.
\]
Working with $i$-complex geometry,  we have only the barred imaginary 
unit $1\mid i$, and so this 
possibility is avoided.

However, we can have recourse to a ``trick''. The action of the 
standard $\gamma^0$-matrix~\cite{ZUB} 
on the complex spinor $\psi \in C^4$ is
\[ \gamma^0 \psi = \left( \begin{array}{cccc} 
1 & 0 & 0 & 0 \\
0 & 1 & 0 & 0 \\
0 & 0 & $-$1 & 0 \\
0 & 0 & 0 & $-$1 
   \end{array} \right) 
\left( \begin{array}{c} \psi_1\\ \psi_2 \\ \psi_3 \\ \psi_4
   \end{array} \right) =
\left( \begin{array}{c} \psi_1\\ \psi_2 \\ $-$\psi_3 \\ $-$\psi_4
   \end{array} \right) ~.
\]
In terms of complexified quaternions we have to find an operation which performs the following transformation
\[
\Psi \equiv \psi_1 + j \psi_2 + \bbox{\iota} (\psi_3 + j \psi_4)
\, \rightarrow \,  \psi_1 + j \psi_2 - \bbox{\iota} (\psi_3 + j \psi_4)~.
\]
The solution is now obvious. The required operation is the 
$\bullet$-involution,
$\Psi \rightarrow \Psi^{\bullet}$. Finally, the Dirac's equation 
\[ \left( \partial_t + \gamma^0 \vec{\gamma} \cdot \vec{\partial} \right)
\Psi (x) \, i
= m \gamma^0 \Psi(x)~,
\]
reads
\begin{equation}
\label{de}
\left( \partial_t  +\bbox{\iota} \vec{h} \cdot \vec{\partial} \right) \Psi(x)
\, i = m \Psi^{\bullet}(x)~.
\end{equation}
Eq.~(\ref{de}) can be concisely rewritten in the following way
\begin{equation}
\label{dec}
D \Psi(x)  = m \Psi^{\bullet}(x)~,
\end{equation}
where
\[ D \equiv \left( \partial_t  +\bbox{\iota} \vec{h} \cdot 
\vec{\partial} \right) \mid i ~.\]

We can immediately check if this equation reduces to the Klein-Gordon equation.
We multiply Eq.~(\ref{dec}) on the left by the barred operator
\[ 
D^{\bullet} \equiv 
\left( \partial_t  - \bbox{\iota} \vec{h} \cdot \vec{\partial} \right) \mid i~,
\]
obtaining
\begin{equation}
\label{mde}
D^{\bullet} D \Psi(x) = 
- \left( \partial_t^2  - \vec{\partial}^{\, 2} \right) \Psi(x) = 
m D^{\bullet} \Psi^{\bullet}(x)~.
\end{equation}
Note that the $\bullet$ involution changes the Dirac equation as follows
\[
D \Psi(x)  = m \Psi^{\bullet}(x) ~~~\rightarrow~~~
D^{\bullet} \Psi^{\bullet}(x)  = m \Psi(x)~,
\]
so Eq.~(\ref{mde}) gives the required Klein-Gordon equation
\[ \left( \partial_{\mu} \partial^{\mu} + m^2 \right) \Psi(x) = 0 ~. \]
If $\Psi(x) \sim e^{-ipx}$, we obtain then from Dirac equation the usual 
Einstein's energy-momentum relation
\[ E^2 = m^2 + \vec{p}^{\, 2}~. \]
It is obvious from the previous discussion that it is not important to pick 
a particular set of quaternionic imaginary units, since the solutions to the 
Dirac equation are completely specified by the anti-commutation relations in 
$\vec{h}$. However, explicit representations can sometimes be helpful in 
making calculations. In the following we shall use $\vec{h}\equiv (i,j,k)$.

In terms of $i$-complex functions which characterize our Dirac spinor
\[ 
\Psi = \psi_1 + j \psi_2 + \bbox{\iota} (\psi_3 + j \psi_4)~,
\]
the Dirac equation can be rewritten as four $i$-complex equations. Instead of
solving these four coupled equations directly, let us try solutions in which 
all four $i$-complex function components from $\psi_1$ to $\psi_4$ share a 
common exponential factor similar to the Klein-Gordon plane wave function
\[ \Psi = \psi_{\vec{p}} \, \,  e^{-ipx}~.\]
Inserting this function into Eq.~(\ref{de}), we obtain
\begin{equation}
\label{de2}
\left( E - \bbox{\iota} \vec{h} \cdot \vec{p} \right) \psi_{\vec{p}}
= m \psi^{\bullet}_{\vec{p}}~.
\end{equation}

Let us first solve this equation in the rest frame of the particle, in which
Eq.~(\ref{de2}) reduces to 
\[ E \psi_{\vec{0}} = m \psi^{\bullet}_{\vec{0}}~. \]
If we pose
\[ \psi_{\vec{p}} = u_{\vec{p}} + v_{\vec{p}}~, \]
where
\[ u_{\vec{p}} \in {\cal H} ~~~~~\mbox{and}~~~~~            
   v_{\vec{p}} \in \bbox{\iota}{\cal H}~,\]
we find the following solutions to the previous equation 
\[ 
u^{(1)}_{\vec{0}}\sim 1 \, ,~~~u^{(2)}_{\vec{0}}\sim j ~~~(E=m);~~~~~
v^{(1)}_{\vec{0}}\sim \bbox{\iota} \, ,~~~
v^{(2)}_{\vec{0}}\sim \bbox{\iota} j ~~~(E=-m)
~.\]
Each one of the spinors $u_{\vec{0}}$ and $v_{\vec{0}}$ has two independent 
solutions. Analogous to
the interpretation of the two-component Pauli spinors, the two independent 
solutions for each one of the spinors will be 
interpreted as the two spin states of a 
spin $\frac{1}{2}$ particle.

For the general case in which the particle is in motion the solutions to the
Dirac equation are obtained as follows:
\[ 
\left( \partial_t  +\bbox{\iota} \vec{h} \cdot \vec{\partial} \right) 
 (u_{\vec{p}} + v_{\vec{p}}) \, e^{-ipx}\, i = 
m (u_{\vec{p}} - v_{\vec{p}} ) \, e^{-ipx}~,
\]
and so
\[ 
\left( E  - \bbox{\iota} \vec{h} \cdot \vec{p} \right) 
 (u_{\vec{p}} + v_{\vec{p}}) = 
m (u_{\vec{p}} - v_{\vec{p}} ) ~.
\]
From the following two coupled equations:
\begin{eqnarray*}
E u_{\vec{p}}  - \bbox{\iota} \vec{h} \cdot \vec{p} \, v_{\vec{p}} & = & 
m u_{\vec{p}}~,\\
E v_{\vec{p}}  - \bbox{\iota} \vec{h} \cdot \vec{p} \, u_{\vec{p}} & = & 
- m v_{\vec{p}}~,
\end{eqnarray*}
we immediately find the {\em desired} complexified quaternionic solutions  to 
the Dirac equation:
\begin{center}
$\sqrt{\frac{|E|+m}{2}}~~\times~~$
\begin{tabular}{|llc|}\hline \hline
 & & \\
~~$1 + \frac{\bbox{\iota} \vec{h} \cdot \vec{p}}{|E|+m}$~, &
~~~$\left( 1 + \frac{\bbox{\iota} \vec{h} \cdot \vec{p}}{|E|+m} \right) \, j$
~,
 & ~~~$E>0$~,~~~\\
 & & \\
\hline
 & & \\
~$\left( 1 - \frac{\bbox{\iota} \vec{h} \cdot \vec{p}}{|E|+m} \right) \, 
\bbox{\iota}$~, &
~~~$\left( 1 - \frac{\bbox{\iota} \vec{h} \cdot \vec{p}}{|E|+m} \right) \, 
\bbox{\iota} j$~,
 & ~~~$E<0$~.~~~\\
 & & \\
\hline \hline
\end{tabular}
\end{center}
The normalization is chosen so that  
\begin{eqnarray*} 
\left( \psi^{\dag}_{\vec{p}} \psi^{\bullet}_{\vec{p}} \right)^{E>0} =
\left( \psi^{\star}_{\vec{p}} \psi_{\vec{p}} \right)^{E>0} 
& ~~=~~ & m~,\\
\left( \psi^{\dag}_{\vec{p}} \psi^{\bullet}_{\vec{p}} \right)^{E<0} =
\left( \psi^{\star}_{\vec{p}} \psi_{\vec{p}} \right)^{E<0} 
& ~~=~~ & - m~,\\
\end{eqnarray*}
or equivalently
\[ \left( \psi^{\dag}_{\vec{p}} \psi_{\vec{p}} \right)_{(1,i)} =
|E|~.
\]
The orthogonality of our solutions is guaranteed by the $i$-complex projection
of inner products.

\subsection{Schr\"odinger-Pauli's equation}
\label{s41}

Let us determine how our complexified quaternionic Dirac equation reduced 
to the Schr\"odinger-Pauli's equation of the electron. We consider the case 
of an electron in the presence of a time-independent electromagnetic field.
Under the assumption of ``minimal electromagnetic coupling'', we replace
\[ 
\partial_{\mu} \mid i \rightarrow \partial_{\mu} \mid i - e A_{\mu}~,
\]
in the Dirac equation to obtain
\[ 
\left( \partial_t \mid i + \gamma^0 \vec{\gamma} \cdot \vec{\partial} \mid i 
- e \gamma^{0} A_{0} + e \gamma^0 \vec{\gamma} \cdot \vec{A}
\right)
\Psi (x) 
= m \gamma^0 \Psi(x)~,
\]
or {\em equivalently}
\[ 
\left[ \partial_t \mid i + \bbox{\iota} \vec{h} \cdot 
\left( \vec{\partial} \mid i + e \vec{A} \right) 
\right]
\Psi (x) 
= \left( m +e A_0 \right) \Psi^{\bullet}(x)~.
\]
In the following we shall use $\vec{p}$ to indicate the ``momentum operator''.
The Dirac equation reads
\[ 
\left[ E - \bbox{\iota} \vec{h} \cdot 
\left( \vec{p} - e \vec{A} \right) 
\right]
\Psi (x) 
= \left( m +e A_0 \right) \Psi^{\bullet}(x)~.
\]
From the previous equation we can write down two coupled equations
\begin{eqnarray*} 
\left( E - m - e A_0 \right) u_{\vec{p}} & = &   
\bbox{\iota} \vec{h} \cdot 
\left( \vec{p} - e \vec{A} \right) 
v_{\vec{p}}~,\\
\left( E + m + e A_0 \right) v_{\vec{p}} & = &   
\bbox{\iota} \vec{h} \cdot 
\left( \vec{p} - e \vec{A} \right) 
u_{\vec{p}}~.
\end{eqnarray*}
For $E=|E|$, we find
\[ 
\left( |E| - m - e A_0 \right) u_{\vec{p}}  =    
\frac{\left[ \bbox{\iota} \vec{h} \cdot 
\left( \vec{p} - e \vec{A} \right) \right] ^2}{|E| + m + e A_0} \, 
u_{\vec{p}}~.
\]
Now, noting that
\[ \partial_x \left( A_y u_{\vec{p}} \right) - 
   A_y \partial_x u_{\vec{p}} =
   \left( \partial_x A_y \right) u_{\vec{p}}~, \]
we find, for $|E| \sim m$ and $A_0 \ll m$,
\[
\left( |E| - m \right) u_{\vec{p}} \sim     
\left[ eA_0 + \frac{1}{2m} \, \left( \vec{p} - e \vec{A} \right)^2
- \frac{e}{2m} \, \bbox{\iota} \vec{h} \cdot \vec{B} \right]
u_{\vec{p}}~.
\]
We recognize the ``Hamiltonian associated with the kinetic 
energy $|E|-m$ of the electron'' characterizing the Schr\"odinger-Pauli's  
equation.

We conclude this section with some considerations about 
the Schr\"odinger's equation
\[ \partial_t \Psi i = \frac{\vec{p}^{\, 2}}{2m} \Psi~.\]
Such an equation assumes the same form both in complex and real/complexified 
quaternionic Quantum Mechanics. Nevertheless,  in the complex world it has 
only {\em one} (complex) solution,  
in the real quaternionic world {\em two} (complex orthogonal) solutions 
and this suggests a their possible identification with the two spin states:
up and down. Finally in the complexified quaternionic world we find 
the ``full'' solution, spin up/down and particle/anti-particle solution.

\subsection{Relativistic Covariance}
\label{s42}

Before to follow the standard approach to the relativistic covariance of
the Dirac's equation, let
us briefly analyze the complexified quaternionic Lorentz transformations.
We can identify using standard idea of affine geometry the coordinates of 
events of Minkowsky space-time, as 
the four-vector $\left(t,\vec{x} \right)$, which by it turns can be 
represented by the complexified quaternion
\[ {\cal X} = t + \bbox{\iota} \vec{h} \cdot \vec{x}~. \]
The Lorentz square of the complex quaternionic position is then
\[  {\cal X}^{\bullet} {\cal X} = 
\left (t - \bbox{\iota} \vec{h} \cdot \vec{x} \right)
\left (t + \bbox{\iota} \vec{h} \cdot \vec{x} \right)
= t^2 - \vec{x}^2~,\]
which represents the translation by complexified quaternions of the 
standard invariant
\[ x^{\mu} x_{\mu} = g_{\mu \nu} x^{\nu} x^{\mu} = t^2 - \vec{x}^{2}~, \]
thanks to the identification 
\[ x^{\mu} ~\leftrightarrow~{\cal X}~,~~~~~
   x_{\mu} ~\leftrightarrow~{\cal X}^{\bullet}~.\]
The Lorentz transformations are concisely described by 
\[ {\cal X}' = \Lambda {\cal X} \Lambda^{\dag}~,
   ~~~~~\Lambda^{\star} \Lambda = 1~,
   ~~~~~\Lambda \in {\cal H}_c~.
\]
Let us introduce the operator
\begin{equation}
{\cal D} \equiv \partial_t - \bbox{\iota} \vec{h} \cdot \vec{\partial}~,
\end{equation}
which represents the quaternionic counterpart of
\[ \partial^{\mu} \equiv \left( \partial_t, - \vec{\partial} \right)~, \]
and which transforms like $\cal X$
\[ {\cal D}' = \Lambda {\cal D} \Lambda^{\dag}~.\]
In order to obtain the relativistic covariance of the Dirac equation  
we must assume that, under 
Lorentz transformations, ${\cal X} \rightarrow {\cal X}'$, 
there is a linear relation between the wave function $\Psi$ in the first 
frame and the wave function $\Psi'$ in the transformed frame, namely
\begin{equation}
\Psi' = {\cal T}(\Lambda) \Psi~.
\end{equation}
Both the wave functions, $\Psi$ and $\Psi'$, 
must satisfy the Dirac equation:
\[ D \Psi = m \Psi^{\bullet}~,~~~~~
   D' \Psi' = m \Psi'^{\bullet}~,
\]
which in terms of the operator $\cal D$ become
\[ {\cal D}^{\bullet} \Psi i = m \Psi^{\bullet}~,~~~~~
   {\cal D}'^{\bullet} \Psi' i = m \Psi'^{\bullet}~,
\]
note that
\[ D \equiv {\cal D}^{\bullet} \mid i~. \]
By observing that under Lorentz transformations the 
${\cal D}^{\bullet}$-operator transforms in the following way
\[ {\cal D}'^{\bullet} = \Lambda^{\bullet} {\cal D}^{\bullet} 
\Lambda^{\star}~,\]
we find for the ``transformed'' Dirac equation
\[ D' \Psi' = 
   \Lambda^{\bullet} {\cal D}^{\bullet} \Lambda^{\star} 
   {\cal T}(\Lambda) \Psi i =
   m \left( {\cal T}(\Lambda) \Psi \right)^{\bullet} =
   m \Psi'~.
\]
After simple algebraic manipulations we obtain
\begin{equation}
{\cal T}(\Lambda) = \Lambda~. 
\end{equation}
A finite transformation is of the form
\[ \exp \left( \vec{h} \cdot \vec{c} \right)
   ~~~~~~~\vec{c} \in {\cal C}(1,\bbox{\iota})~. \]
For spatial rotation $\cal T$ is unitary (generators $\vec{h}$), 
whereas it is hermitian for Lorentz boosts 
(generators $\bbox{\iota} \vec{h}$). 
It is immediate to observe that $\Psi \Psi^{\dag}$ transforms
as the four-dimensional vector $\cal X$
\begin{equation}
\Psi' \Psi'^{\dag} = \Lambda \Psi \Psi^{\dag} \Lambda^{\dag}~,
\end{equation}
whereas $\Psi^{\star} \Psi$ transforms like a scalar
\begin{equation}
\Psi'^{\star} \Psi' = \Psi^{\star} \Lambda^{\star} \Lambda \Psi =
                      \Psi^{\star} \Psi~.
\end{equation}
An explicit calculation for the Dirac's spinors gives
\begin{eqnarray*} 
\Psi_{\vec{p}} \Psi_{\vec{p}}^{\dag} & ~=~ & 
|E| \pm \bbox{\iota} \vec{h} \cdot \vec{p}~,\\
\Psi_{\vec{p}}^{\star} \Psi_{\vec{p}} & ~=~ & 
\pm \, m~.
\end{eqnarray*}
We will show in the next section that the parity operation is expressed 
by the $\bullet$-involution, so we observe that the multiplication
by $\bbox{\iota}$ transform scalars and vectors in pseudo-scalars and
pseudo-vectors:
\begin{eqnarray*}
\Psi^{\star} \Psi & ~~~\mbox{scalar}~,\\ 
\Psi \Psi^{\dag}  & ~~~\mbox{vector}~,\\
\bbox{\iota} \Psi^{\star} \Psi & ~~~\mbox{pseudo-scalar}~,\\ 
\bbox{\iota} \Psi \Psi^{\dag}  & ~~~\mbox{pseudo-vector}~.
\end{eqnarray*}

\subsection{Spin Operator}
\label{s43}

We conclude this section by giving the explicit form of the spin operator. 
We know that the spin operator is related to space rotations, thus by 
considering an infinitesimal rotation around $x$, and finding the 
corresponding transformation of the wave function $\Psi$, we obtain
\begin{equation}
{\cal S}_x = - \frac{i \mid i}{2}~.
\end{equation}
Thus, the four solutions $u^{1,2}_{\vec{p}}$, correspond to positive 
energy solutions with ${\cal S}=\frac{1}{2}$ and for 
$\vec{p}\equiv (p_x,0,0)$ to 
${\cal S}_x = \frac{1}{2}, -\frac{1}{2}, \frac{1}{2}, -\frac{1}{2}$
respectively. Our polarization direction is the $x$-axis because the 
imaginary unit $i$ has been associated with $p_x$.


\section{CPT operation}
\label{s5}

In this section we discuss the CPT operation. We will show that in the 
complexified quaternionic world, it assumes a simple form and represents
a mapping of our spinors $\Psi$ in their dual space. In order to simply
the mathematical language, we shall use the following notation
\begin{eqnarray*}
\Psi   & ~\equiv~ & \Psi (x)~,\\
\Psi_P & ~\equiv~ & \Psi'(-\vec{x},t)~,\\
\Psi_C & ~\equiv~ & \Psi_C (x)~,\\
\Psi_T & ~\equiv~ & \Psi'(\vec{x},-t)~.
\end{eqnarray*}

\subsection{Parity}
\label{s51}

We start from the complexified quaternionic Dirac's equation
\[ \left( \partial_t + \bbox{\iota} \vec{h} \cdot \vec{\partial} \right) 
\Psi i 
= m \Psi^{\bullet}~, \] 
and we perform the required coordinates transformation (space inversion)
\[ \vec{x} ~~\rightarrow ~~ -\vec{x}~.\]
We obtain the transformed Dirac equation:
\begin{equation}
\label{pe}
\left( \partial_t - \bbox{\iota} \vec{h} \cdot \vec{\partial}
\right) \Psi_P \, i = 
m \Psi_P^{\bullet}~. 
\end{equation} 
In our formalism is now very easy to find the relation between the transformed 
wave function, $\Psi_P$, and the wave function in the first frame, $\Psi$.
The $\bullet$-involution modifies the Dirac equation as follows 
\[ \left( \partial_t - \bbox{\iota} \vec{h} \cdot \vec{\partial}
\right) \Psi^{\bullet} i = 
m \Psi~, \] 
and so by comparison of 
this equation with Eq.~(\ref{pe}), we immediately find
\begin{center}
\begin{tabular}{|c|} \hline \\ 
~~$\Psi_P \equiv \Psi^{\bullet}$~~~\\  \\
\hline
\end{tabular}
\end{center}
And as anticipated in the previous section, the parity operation is expressed
by the $\bullet$-involution.

\subsection{Charge Conjugation}
\label{s52}

To discuss charge conjugation, we introduce the potential 
$\left(A^0,\vec{A} \right)$ by performing the following change in our Dirac 
equation
\begin{eqnarray*}
\partial_t \mid i & ~\rightarrow~ & \partial_t \mid i + e A^0~,\\
\vec{\partial} \mid i & ~\rightarrow~ & \vec{\partial} \mid i - e \vec{A}~.
\end{eqnarray*}
The ``modified'' Dirac equation now reads
\[ \left[ \partial_t \mid i + e A^0 + \bbox{\iota} \vec{h} \cdot 
   \left( \vec{\partial} \mid i - e \vec{A} \right) \right]
    \Psi  = m \Psi^{\bullet}~. \] 
The charge conjugation requires the change $e \rightarrow -e$
\begin{equation}
\label{ce}
 \left[ \partial_t \mid i - e A^0 + \bbox{\iota} \vec{h} \cdot 
 \left( \vec{\partial} \mid i + e \vec{A} \right) \right]
 \Psi_C  = m \Psi_C^{\bullet}~.
\end{equation}
By multiplying the Dirac's equation by $\bbox{\iota}$, 
\[ \left[ \partial_t \mid i + e A^0 + \bbox{\iota} \vec{h} \cdot 
   \left( \vec{\partial} \mid i - e \vec{A} \right) \right]
   \left( \bbox{\iota} \Psi \right)  = 
    - m \left( \bbox{\iota} \Psi \right)^{\bullet}~, \] 
and by $j$ on the right, we find 
\[ \left[ -\partial_t \mid i + e A^0 + \bbox{\iota} \vec{h} \cdot 
   \left( -\vec{\partial} \mid i - e \vec{A} \right) \right]
   \left( \bbox{\iota} \Psi j \right)  = 
    - m \left( \bbox{\iota} \Psi j \right)^{\bullet}~. \]
The last equation when rewritten as
\[ \left[ \partial_t \mid i - e A^0 + \bbox{\iota} \vec{h} \cdot 
   \left( \vec{\partial} \mid i + e \vec{A} \right) \right]
   \left( \bbox{\iota} \Psi j \right)  = 
    m \left( \bbox{\iota} \Psi j \right)^{\bullet}~, \]
and confronted with~(\ref{ce}) gives
\begin{center}
\begin{tabular}{|c|} \hline \\ 
~~$\Psi_C \equiv \bbox{\iota} \Psi j$~~\\  \\
\hline
\end{tabular}
\end{center}
Thus, the charge conjugation is expressed in the complexified quaternionic 
formalism, by the multiplication by imaginary units, 
$\bbox{\iota}$ (mapping in the dual space) and $j$ (spin flip).

\subsection{Time Reversal}
\label{s53}

By noting that the time reversal requires
\[ A^0 \rightarrow A^0~,~~~~~\vec{A} \rightarrow -\vec{A}~,\]
we have
\begin{equation}
\label{te}
 \left[ -\partial_t \mid i + e A^0 + \bbox{\iota} \vec{h} \cdot 
 \left( \vec{\partial} \mid i + e \vec{A} \right) \right]
 \Psi_T  = m \Psi_T^{\bullet}~.
\end{equation}
Let us multiply, from the right, the Dirac's equation by $j$ 
\[ \left[ -\partial_t \mid i + e A^0 + \bbox{\iota} \vec{h} \cdot 
   \left( -\vec{\partial} \mid i - e \vec{A} \right) \right]
   \left( \Psi j \right)  = 
    m \left( \Psi j \right)^{\bullet}~. \]
The $\bullet$-involution modifies the previous equation as follows  
\[ \left[ -\partial_t \mid i + e A^0 + \bbox{\iota} \vec{h} \cdot 
   \left( \vec{\partial} \mid i + e \vec{A} \right) \right]
   \left( \Psi j \right)^{\bullet}  = 
    m \left( \Psi j \right)~. \]
By  comparing with~(\ref{te}), we find 
\begin{center}
\begin{tabular}{|c|} \hline \\ 
~~$\Psi_T \equiv \Psi^{\bullet} j$~~\\  \\
\hline
\end{tabular}
\end{center}

It is now immediate to obtain the ``full'' CPT operation
\begin{center}
\begin{tabular}{|c|} \hline \hline \\ 
~~$\Psi_{CPT}(x') \equiv \bbox{\iota} \Psi (x) e^{i\phi}$~~\\  \\
\hline \hline 
\end{tabular}
\end{center}

We conclude this section with some considerations about the geometric  
interpretations of the complexified quaternionic imaginary units.
The {\em pure} quaternionic imaginary units, $\vec{h}$, represent 
the generators of the space rotations, the complexified quaternionic 
products, $\bbox{\iota}\vec{h}$, are related to the boost generators and
finally the {\em pure} complex imaginary unit, $\bbox{\iota}$, gives 
rotations in the plane individuated by $\Psi$ and its dual image 
$\bbox{\iota} \Psi$.


\section{Conclusions}
\label{s6}

We conclude this paper by showing a surprising possibility of translation
between standard (complex) Quantum Mechanics and complexified quaternionic QM
with $\bbox{\iota}$-complex geometry.

We begin by recalling the ``symplectic'' quaternionic representation of a 
complexified quaternionic (state) $q_c$
\[ q_c = q_1 + \bbox{\iota} q _2 ~~~~~q_{1,2} \in {\cal H}~,\]
by the quaternionic column vector
\begin{eqnarray*}
q_c & \leftrightarrow & 
\left( \begin{array}{c} q_1 \\ q_2 \end{array} \right)~.
\end{eqnarray*}
We now identify the operator representation of $\bbox{\iota}$ consistent 
with the above identification:
\begin{eqnarray*}
\bbox{\iota} & \leftrightarrow & 
\left( \begin{array}{cc} 0 & $-$1 \\ 1 & 0  \end{array} \right)~.
\end{eqnarray*}
In order to obtain a translation between $2\times 2$ real quaternionic 
(barred) matrices
\[
\left( \begin{array}{cc} 
q_1 + q_2 \mid i & ~~p_1 + p_2 \mid i \\
r_1 + r_2 \mid i & ~~s_1 + s_2 \mid i 
\end{array} \right)~,
\]
and barred complexified quaternions
\[ q_c + p_c \mid  i~, \]
we need to obtain the complexified quaternionic counterpart of
\[
\left( \begin{array}{cc} 
1 & 0 \\
0 & $-$1 
\end{array} \right)~.
\]
This is soon achieved by the $\bullet$-involution. Thus, we have the 
following set of rules for the required translation
\[
1 ~ \leftrightarrow ~  
\left( \begin{array}{cc} 1 & 0 \\ 0 & 1  \end{array} \right)~,~~~
\bbox{\iota} ~ \leftrightarrow ~  
\left( \begin{array}{cc} 0 & $-$1 \\ 1 & 0  \end{array} \right)~,~~~
\bullet \mbox{-inv} ~ \leftrightarrow ~  
\left( \begin{array}{cc} 1 & 0 \\ 0 & $-$1  \end{array} \right)~,~~~
\bbox{\iota} \times \bullet \mbox{-inv}  ~ \leftrightarrow ~  
\left( \begin{array}{cc} 0 & 1 \\ 1 & 0  \end{array} \right)~.
\]
The basis is
\[ 1~,~~1\mid i~,~~\vec{h}~,~~\vec{h}\mid i~,\]
and so we restore the needed 32 real parameters. Since $2\times 2$ real 
quaternionic (barred) matrices are related to $4\times 4$ complex 
matrices~\cite{T}, we can immediately obtain the translation between 
4-dimensional complex matrices and one-dimensional complexified operators.

In conclusion we have completed our previous work on the possibility to
formulated a consistent Quantum Mechanics by noncommutative fields~\cite{DR},
by discussing the main features of a complexified quaternionic approach based 
on $\bbox{\iota}$-complex geometry. We hope that the complexified 
quaternionic Dirac's equation elaborated in this article and the translation 
given in this section demonstrate the possible potentialities in the use of
noncommutative numerical fields (and in particular complexified quaternions)
in formulating physical theories.

Nevertheless, we wish to insist upon the non-complete nature of the 
translation and hence the non-triviality in the choice to adopt 
complexified quaternions as underlying numerical field. 
Many geometric interpretations, hidden in the ``complex world'', 
can be pointed out by the complexified quaternionic algebra~\cite{DR2}


\end{document}